\documentclass{aa}
\input{psfig}
\begin{document}
 
\thesaurus{04(02.08.1; 02.09.1; 09.11.1; 11.05.2; 11.11.1; 11.19.2)}
\title{Global Spiral Modes in Star-Forming Gravitating Disks}
\author{Vladimir Korchagin \inst{1} \thanks{\emph{Present address:} 
  National Astronomical Observatory, Osawa 2-21-1, Mitaka, Tokyo 181, Japan}
  and Christian Theis \inst{2} }
  \institute{Institute of Physics, Stachki 194, Rostov-on-Don, Russia,
  email: vik@rsuss1.rnd.runnet.ru \and
  Institut f\"ur Theoretische Physik und Astrophysik,
  D--24098 Kiel, Germany,  email: theis@astrophysik.uni-kiel.de}
\offprints{Ch.\ Theis}
\date{received \today; accepted}
\maketitle



\begin{abstract}
  Using 2D nonlinear simulations, we study the generation and 
  nonlinear evolution of spiral structure in a star-forming 
  multi-component gravitating disk. 
  We confirm in agreement with previous studies 
  the destabilizing role of a cold gaseous component
  and extend this conclusion for multi-component star-forming disks
  exchanging mass and momentum between its components.
  We show that the spiral structure growing on a non-stationary multi-phase
  background reaches its saturation in a similar manner like the 
  one-component disks. The spiral structure survives even if
  most of the gas is transformed into stellar remnants 
  of larger velocity dispersion.
\end{abstract}


\section{Introduction}

The regular spiral arms observed in nearby disk 
galaxies are a manifestation of global spiral modes developed in galactic 
disks. Numerous linear and nonlinear studies have demonstrated that  
self-gravity plays a major role in the generation of global spiral modes
(see, e.g.\ Binney and Tremaine \cite{binney87}). 

Growth of spirals on early stages of 
galactic evolution, however, occurs on a rapidly changing background. 
The precursors of present systems look bluer than old galaxies 
giving evidence for an enhanced star formation rate in early epochs.
The colors and irregular structure seen in
intermediate red-shifted galaxies suggest that they are a population of
galaxies with massive star
formation rapidly consuming gas in the galactic disks. 
These galaxies are the systems where star-formation might
considerably change the physics of spiral dynamics. 

In this paper we address the question how the evolution of
the large-scale spiral perturbations is affected by the 
processes of mass and momentum
exchange in a star forming disk. Kato (\cite{kato72}, \cite{kato74}) 
discussed the effects of star-gas mass exchange  
in a linearly growing spiral mode. Using the WKB-approximation he found, that
under some conditions the mass exchange between stars and gas
can excite density waves. We extend this study by following the dynamics
of the nonlinear hydrodynamical equations for three interacting
components: the massive stars, the low mass stars (or stellar remnants)
and the gas. We perform a series of 2D multi-phase hydrodynamical simulations
of the dynamics of unstable multi-component disks, and compare our results
with the morphological properties of spirals, growing in one-component
self-gravitating disks. As a particular realization of 
the mass transformations in star-forming systems
we choose the description of K\"oppen et al. (\cite{koeppen95}) which is
a subset of the more detailed ''chemo-dynamical'' models developed by
Theis et al.\ (\cite{theis92}), Hensler et al.\ (\cite{hensler93}), 
Samland (\cite{samland94}) and Samland et al.\ (\cite{samland97}). 
Apart from previous papers investigating the interaction network of a 
multi-component system for spherical (Theis et al.\ \cite{theis92}) or 
axisymmetric galaxies (Samland et al.\ \cite{samland97}) we deal here 
with the evolution of thin, but non-axisymmetric systems. This allows 
us to study the influence of the interactions between the components
on the evolution of spiral structure.


\section{Basic Equations for a Three-Component Star Forming Disk}

The physical processes in star forming galactic disks are too
complicated for a detailed analysis from first principles.
Among numerous dynamical and interchange
processes one has to choose those which have a major influence on the
evolution of the system.
We perform 
our analysis using a simplified model which splits the disk into three 
components regulated by time-dependent mass transformations. We take into
account a gas component containing all phases of the interstellar medium,
and two stellar components, the massive and the low mass stars. 
The less massive stars are
assumed to have no influence on the interstellar medium, and are simply
accumulated as remnants, whereas massive stars are assumed to be responsible
for gas heating.

The chain of mass transformations within the
model includes spontaneous 
star formation and ejection of the stellar mass back into the gas
phase. 
For the description of the star formation rate we use the 
approach developed by K\"oppen et al.\ (\cite{koeppen95}) which is a 
basic skeleton of the interaction scheme used in chemo-dynamical models
(e.g.\ Theis et al.\ \cite{theis92}, Samland et al.\ \cite{samland97}). 
It describes the
spontaneous stellar birth as a power law function of the gas density
with an efficiency depending on the temperature of 
the gas. The latter just has to guarantee that the efficiency drops down
with increasing temperature. In this scenario, the chain of mass 
transformation processes can be described by the following set of equations:
\begin{equation}
 {dg \over dt} = - \Psi(g,T) + \eta s/ \tau
  \label{eq1}
\end{equation}
\begin{equation}
 {ds \over dt} = \zeta \Psi(g,T) - s/\tau
 \label{eq2}
\end{equation}
\begin{equation}
 {dr \over dt} = (1-\zeta)\Psi(g,T) + (1-\eta) s/ \tau
 \label{eq3}
\end{equation}
where the star formation rate $\Psi(g,T)$ is given by a power law dependence
on the gas density $g$ and a function of the gas temperature $T$ 
\begin{equation}
  \Psi(g,T) = C_n f(T) g^n \mbox{.}
   \label{eq4}
\end{equation}
$s$ denotes the density of massive stars, whereas $r$ corresponds to the
low-mass stars and the stellar remnants.
The parameters $\zeta$ and $\eta$ determine the fraction of newly born 
massive stars, and the fraction of gas ejected by massive stars into the ISM,
respectively. The parameter $\tau$ corresponds to the mean stellar lifetime
of a massive star.

In general, for the determination of the ''efficiency'' factor $f(T)$
which depends on the gas temperature,
one has to integrate the energy balance equation. K\"oppen 
et al.\ (\cite{koeppen95}) noticed, however, that the model can be simplified,
if the time-scales for heating and cooling of the gas are shorter than
the characteristic dynamical time-scale of the system. In that case,
the effective star formation rate depends only weakly on the exponent $n$
and the efficiency function $f(T)$, and the star formation effectively follows
the Schmidt law with the 'true' star formation rate depending quadratically
on the gas density. Additionally, due to the self-regulation inherent in
this star formation prescription, the effective star formation rate is not
strongly affected by substantial variations of $f$ or $C_n$.
Therefore, we can restrict our model to the 
simpler expression
\begin{equation}
   \Psi = C_2 g^2 \mbox{.}
  \label{eq5}
\end{equation}
We will use this expression in our analysis.

For the description of the spatial dynamics of the multi-component 
disk we will use a fluid dynamical approach. In this approach, gas, 
stars and remnants are considered as three  
fluids, coupled by nonlinear interchange processes and by the common
gravity. While the application of fluid dynamics for
the description of the gas is quite natural, it is not obvious
that such an approach can be used for the collisionless components of the
disk represented by stars and remnants. Kikuchi et al.\ (\cite{kikuchi97}) 
analyzed this question and found that the stability 
properties of disks obtained in fluid approximation are in good qualitative,
and to some extent in quantitative agreement with the stability properties
of the collisionless models. We will use therefore a fluid approximation
in the analysis of perturbations of a multi-component disk.
The behavior of our model disk is described by the continuity
equations, written for each component, a set of momentum equations, and
the Poisson equation. In cylindrical coordinates they are:

\begin{equation}
   {D_g \sigma_g \over Dt} =
   - C_2 \sigma_g^2 + \eta {\sigma_s \over \tau}
   \label{eq6}
\end{equation}

\begin{equation}
  {D_s \sigma_s \over Dt} =
  \zeta C_2 \sigma_g^2 - {\sigma_s \over \tau}
  \label{eq7}
\end{equation}

\begin{equation}
  {D_r \sigma_r \over Dt} =
  (1-\zeta) C_2 \sigma_g^2 +(1-\eta) {\sigma_s \over \tau}
  \label{eq8}
\end{equation}

\noindent
where $D_{g,s,r} /Dt$ are the corresponding substantial time derivatives
written in cylindrical coordinates:

\begin{equation}
  {D_{g,s,r} \over Dt} = 
  {\partial \over \partial t} + {1 \over r} {\partial \over \partial r}
  r u_{g,s,r}  +
  {1 \over r} {\partial \over \partial \phi} v_{g,s,r} 
  \label{eq9}
\end{equation}

\noindent
Using equations (\ref{eq6}) -- (\ref{eq8}) and definition (\ref{eq9})
the momentum equations can be written as
\begin{eqnarray}
  \label{eq10}
  \sigma_g {D {\bf v_g} \over Dt} + \nabla P_g + \sigma_g \nabla \Big(
  \Phi + \Phi_H + \Phi_B \Big ) = \nonumber \\
    - C_2 \sigma_g^2{\bf v_g} + \eta{\sigma_s \over \tau}{\bf v_s}
\end{eqnarray}

\begin{eqnarray}
  \label{eq11}
  \sigma_s {D {\bf v_s} \over Dt} + \nabla P_s + \sigma_s \nabla \Big(
  \Phi + \Phi_H + \Phi_B \Big ) = \nonumber \\
     \zeta C_2 \sigma_g^2 {\bf v_g} - {\sigma_s \over \tau} {\bf v_s}
\end{eqnarray}

\begin{eqnarray}
  \label{eq12}
  \sigma_r {D {\bf v_r} \over Dt} + \nabla P_r + \sigma_r \nabla \Big(
  \Phi + \Phi_H + \Phi_B \Big ) = \nonumber \\
     (1-\zeta) C_2 \sigma_g^2{\bf v_g}+(1-\eta) {\sigma_s \over \tau} {\bf v_s}
\end{eqnarray}

\noindent
Here $\sigma_{g,s,r}$ are the surface densities, and $u_{g,s,r}$ and
$v_{g,s,r}$ are the radial and azimuthal components of the 
velocities  ${\bf v_{g,s,r}}$ of gas, stars and
remnants in the disk. 

Without the exchange processes, i.e.\ when the right-hand sides of the
equations (\ref{eq10}) -- (\ref{eq12}) are zero,  
the dynamics of these quantities is determined by
the partial "pressures" of the components $ P_{g,s,r}$, the self-gravity
of the disk $\Phi$ and the external gravity of the halo and bulge, $\Phi_H$
and $\Phi_B$. Mass transformations between the components give an additional
factor for the momentum balance and have to be taken explicitly into account
in the numerical simulations.

The gravitational potential $\Phi$ is determined by the overall density of
all components, and can be written in the form of a Poisson integral as

\begin{eqnarray}
  \label{eq13}
  \Phi = 
  -G\int_{R_{\rm in}}^{R_{\rm out}}(\sigma_g(r^{\prime},\phi^{\prime})+
  \sigma_s(r^{\prime},\phi^{\prime})+
  \sigma_r(r^{\prime},\phi^{\prime})) r^{\prime}dr^{\prime} \nonumber \\
    \cdot \int_{0}^{2 \pi}{d \phi^{\prime} \over
    {\sqrt{r^{2}+r^{\prime 2} - 2rr^{\prime} \cos (\phi-\phi^{\prime})}}}
    \hspace*{1cm}
\end{eqnarray}
The equation of state closes the system of equations (\ref{eq6}) --
(\ref{eq12}). Throughout the simulations we use a polytropic equation 
of state applied to all three components:
\begin{equation} 
    P_{g,s,r} = K_{g,s,r} \sigma_{g,s,r}^{\gamma_{g,s,r}}
    \label{eq14}
\end{equation}
Equations (\ref{eq6})-(\ref{eq14}) are used for the analysis 
described in the subsequent sections.  

In our simulations we will use a "galactic" system of units
in which $R=20$kpc, $M=10^{11} M_{\odot}$ and the gravitational constant
$G$ is unity. With this choice we have a unit of time equal to
$1.33\times 10^8$ yr, and a unit of velocity equal to $147$kms$^{-1}$.


\section{Axisymmetric Distributions}

We assume that all three components have axisymmetric 
flat rotation curves in the outer regions of the disk.  
The quasi-stationary equilibrium rotation is jointly supported
by the gravity of the halo and bulge, the self-gravity of the disk and the
pressure gradient:
\begin{equation}
 r\Omega_i^{2}(r)={\partial \over {\partial r}}(\Phi_H + \Phi_B + \Phi_{0i}
 + P_{0i} )
\label{eq15}
\end{equation}
Here $\Omega_i(r)$ is the angular velocity in the disk,
$\Phi_{0i}$ is the axisymmetric self-gravitating potential, 
and $P_{0i}$ are the partial pressures of the components corresponding 
to their unperturbed density distributions.

In our simulations, we consider two types of the equilibrium rotation curves
and surface density distributions. In the first experiment, we choose the
exponentially decreasing surface density of the disk with density distributions
of all three components given by the expression

\begin{eqnarray}
  \label{eq16}
  \sigma_{g,s,r}(r) = M_{g,s,r} \cdot
      \exp\Big(-d_1 \cdot\sqrt{d_2 + d_3 \cdot r^2}\Big) \cdot
      \nonumber \\
  \Big[1-\exp\Big(-{(r-R_{\rm out})^2 \over r_s^2} \Big)\Big]^4
\end{eqnarray}

\noindent
with the normalization constants $M_{g,s,r}$ which are the
masses of the gaseous, stellar and remnant component. 
The halo-bulge potential determining the rotational curve (\ref{eq16}) 
is similar to that used by Vauterin and Dejonghe (\cite{vauterin96}). 
Namely, we assume that the external potential has the form
\begin{equation}
 {\partial \Phi_H \over \partial r} +  {\partial \Phi_B \over \partial r} =
  M_H{ r \over (r^2+R_H^2)^{3\over2} } + M_B{ r \over (r^2+R_B^2)^{3\over2} }
  \mbox{.}
\label{eq17}
\end{equation} 

In a thin disk with sharp boundaries the potential diverges at its edges.
To study to what extent the sharp cut-off in an unperturbed density distribution
influences the results, we performed, additionally, experiments using 
a Gaussian-type density distribution vanishing at both boundaries:
\begin{eqnarray}
  \label{eq18}
  \sigma_{g,s,r}(r) = M_{g,s,r}\Big[1-\exp\Big(-{(r-R_{\rm in})^2 
     \over r_s^2 } \Big) \Big]^4 \cdot \nonumber \\
  \hspace*{0.4cm} \Big[1-\exp\Big(-{(r-R_{\rm out})^2 
     \over r_s^2} \Big)\Big]^4
  \exp\Big(-{(r-R_0)^2 \over \omega} \Big)
\end{eqnarray}

\noindent
In this case the disk is kept in centrifugal equilibrium by an
alternative bulge/halo distribution resulting in a radial acceleration of
the form
\begin{equation}
 {\partial \Phi_H \over \partial r} +  {\partial \Phi_B \over \partial r} =
 v_{\infty}^2 {r \over r^2+R_H^2 } + M_B{ r \over (r^2+R_B^2)^{3\over2} }
  \mbox{.}
\label{eq19}
\end{equation}

\noindent
Both halo and bulge potentials, described by the equations (\ref{eq17}) 
and (\ref{eq19}) dictate the "flat" rotation in the outer region of the disk.
The parameters $R_H$ and $R_B$ in equations (\ref{eq17}) and (\ref{eq19}) 
determine the spatial scales
of the halo and bulge density distributions, $v_{\infty}$ gives the asymptotic 
value of the rotational velocity, and $M_H$ and $M_B$ determine the masses 
of the halo and the central bulge. 


\section{Parameters of the Model}

We consider the dynamics of the disk by setting the radius of the innermost 
boundary cell to $R_{\rm in} = 0.1$, 
and the radius of the outer boundary cell 
to $R_{\rm out}=1$. 

The group of parameters  $\tau, \zeta, \eta$ and $C_2$ in the right-hand side
of the continuity equations (\ref{eq6})-(\ref{eq8}) govern the 
interchange processes between the components. 
Following K\"oppen et al.\ (\cite{koeppen95}) we choose the mean
stellar lifetime $\tau = 10$Myr, or in our units $\tau = 0.075$. The
mass fraction $\zeta$ of the newly formed massive stars was set to 0.12.
This value corresponds to a Salpeter-IMF ranging from 0.1 $M_{\odot}$ to 
100 $M_{\odot}$ and a lower mass limit of massive stars of 10 $M_{\odot}$.
The fraction $\eta$ of mass ejected by massive stars back to the 
interstellar medium was taken to be 0.9.

The parameter $C_2$ was set to 0.1. With this choice, the maximum 
star-formation rate in our model is 0.025, or $\approx 18 M_{\odot}/\mbox{yr}$
if the initial mass of the gaseous disk is equal to 0.5 in our units.
This value of the star-formation rate is close to the maximum star-formation
rate obtained in chemo-dynamical models for the evolution of disk galaxies 
(Samland \cite{samland94}). 

We have assumed that the gas component of the disk is mainly composed
of mono-atomic hydrogen with a volume polytropic index $\gamma_g=1.67$. 
The polytropic constant for the collisionless stellar component and the 
remnants was set to 2.0. There are a few arguments in favor of this choice.
Marochnik (\cite{marochnik66}) found that in a rigidly 
rotating disk the dynamics of
perturbations can be described by introducing the polytropic equation of
state with $\gamma_s = 2$. This value is consistent with the empirical
''square root law'' found by Bottema (\cite{bottema93}) in his studies of 
nearby spiral galaxies. He found, that the surface density distribution of 
stars, and their radial velocity dispersion are related as 
$c_s \propto \sqrt{ \sigma(r)}$. It is easy to see, that such a 
''square root law''
requires the value of the effective polytropic index to be $\gamma_s = 2$.
Kikuchi et al.\ (\cite{kikuchi97}) made a detailed comparison 
of the linear stability properties of the exact collisionless models 
investigated by Vauterin \& Dejonghe (\cite{vauterin96}) with the stability 
properties of this model studied in a fluid dynamical approach. 
They found a full qualitative agreement
between these two approaches. Thus, a fluid dynamical approximation
can be used for the analysis of the multi-component disks. The constant
$K_g$ was set to be 0.04 resulting in a Toomre-stable disk
(for details, see Sect.\ \ref{onecompdisk}). The values $K_s$ and
$K_r$ have been set to twice this value. This choice corresponds to a 
larger ''sound'' velocity of the stars, by this mimicking as well
the dynamical heating of disk stars as the lack of dissipation in the 
stellar component.

All parameters including those which are not discussed in this section 
are listed in the tables in the appendix.


\section{The Code} 

For solving the multi-component hydrodynamical equations (\ref{eq6})
-- (\ref{eq12}) we 
use a second order Van Leer advection scheme as implemented by Stone \&
Norman (\cite{stone92}) in a general purpose fluid dynamics code, 
called ZEUS-2D. This code was designed for modeling astrophysical 
systems in two spatial dimensions, and can be used
for simulations of a wide variety of astrophysical processes. 
The ZEUS-2D code uses sufficiently accurate hydrodynamical
algorithms which allow to add complex physical effects in a self-consistent
fashion. This code provides therefore a good basis for the implementation of
the nonlinear mass transfer processes into the multi-phase hydrodynamics.

The Eulerian codes with the Van Leer advection scheme were successfully
used for the investigation of the stability of self-gravitating disks
(Laughlin \& R\'o\.zyczka \cite{laughlin96}, Laughlin et al.\
\cite{laughlin97}, \cite{laughlin98}). The main
difference between the ''standard'' ZEUS-type codes and our one is the
introduction of mass and momentum interchange processes between
different components. These processes can be computed at the first sub-step
of the ZEUS-type code which makes a generalization of the ZEUS-type
code straightforward.

Briefly, the code
solves the hydrodynamical equations using equally spaced azimuthal
zones and logarithmically spaced radial zones. For the simulations discussed
here we mainly employed a grid with $256\times256$ cells.
To advance the solutions due to interchange processes given by the 
right-hand sides of the equations (\ref{eq6})-(\ref{eq8}) we used
the fifth order Cash-Karp Runge-Kutta routine with the time step limitation
imposed by the Courant-Friedrichs-Levy criterion and the values of the
parameters $\tau$ and $C_2$ in mass and momentum interchange processes.
The Poisson equation is solved by applying the 2D Fourier convolution 
theorem in polar coordinates.  


\section{Global Instability of a One-Component Disk}
\label{onecompdisk}

In order to understand the influence of mass and momentum exchange processes
on the evolution of global modes in a self-gravitating disk we performed 
simulations of the dynamics of a one-component stellar disk.
The equilibrium stellar disk was chosen to have the surface density 
distribution (\ref{eq16}). The parameter $M_s$ for the stellar disk was 
selected so that the disk's total mass is equal to 0.5.  

\begin{figure}[ptbh]
   \psfig{figure=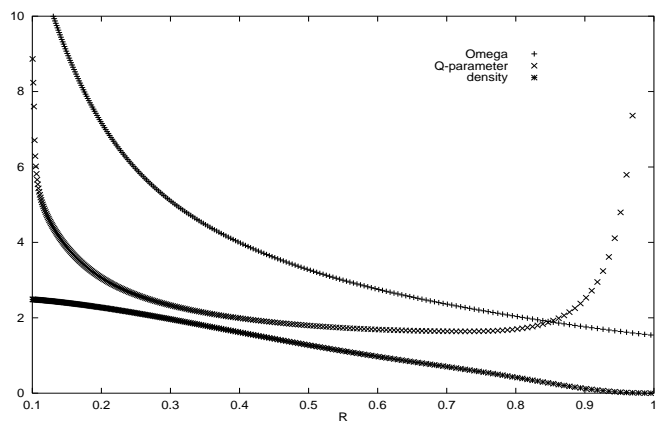,height=5.5cm,width=7.5cm,angle=270,bbllx=70pt,bblly=40pt,bburx=560pt,bbury=650pt}
   \caption[ ]{
          Radial dependence of the angular velocity $\Omega$, 
          the Toomre-parameter $Q$, and the equilibrium density $\sigma$
          for a purely stellar disk ($\gamma=2$) with exponential surface
          density distribution.
       \label{fig1}}
\end{figure}

\begin{figure}[ptbh]
   \psfig{figure=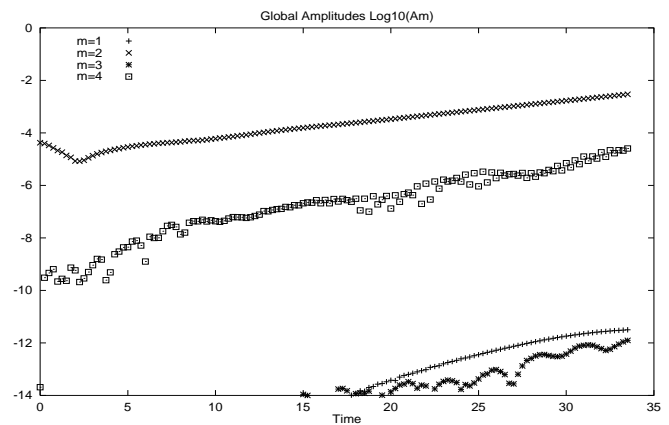,height=5.5cm,width=7.5cm,angle=270,bbllx=70pt,bblly=40pt,bburx=560pt,bbury=650pt}
   \caption[ ]{
       Temporal evolution of the global amplitudes $\log(A_m)$
       ($m=1,2,3,4$) for a purely stellar disk ($\gamma=2$)
        with exponential surface density distribution.
    \label{fig2}}
\end{figure}

\begin{figure*}[ptbh]
   \centerline{
   \psfig{figure=fig3.ps,height=21.cm,width=14.8cm,bbllx=0pt,bblly=0pt,bburx=550pt,bbury=750pt}
   }
   \caption[ ]{
          Contour maps of the radial velocity (in kms$^{-1}$) 
          of a purely stellar disk 
          ($\gamma=2$) with an exponential surface density distribution
          at different times: $t=0$ (upper left), $t=5$ (upper right),
          $t=10,15,20,25$. The contours give 30\%, 50\%, 70\% and 90\% of
          the maximum velocity in each diagram. The dotted lines correspond
          to negative velocities, whereas the solid lines give positive
          velocities. The zero-velocity contour is shown with a dashed
          line.
   \label{fig3}}
\end{figure*}

Figure \ref{fig1} shows the equilibrium properties of the stellar disk
used in the analysis. The Toomre $Q$-parameter, which is defined for the stellar 
disk as 
$Q \equiv c_s\kappa /3.36 G \sigma$, has a profile typical
for the density distribution given by equation (\ref{eq16}).
The $Q$-profile rises towards the boundaries of the disk, with a minimum
value of 1.64 at radius 0.73 indicating a globally stable disk with
respect to Toomre's stability criterion $Q>1$. 

The one-component
stellar disk was perturbed with an $m-$armed perturbation of the form

\begin{eqnarray}
  \label{eq20}
  \sigma_s(r,\phi)  & = & \sigma_s(r) \cdot 
            [1 + 0.001 \cos(m \phi)] \cdot \hspace{3cm} \\
      & &      \cdot \Big[1-\exp\Big(-{(r-R_{\rm in})^2 
                   \over r_s^2 }\Big)\Big]^5 \cdot \nonumber \\
      & &     \Big[1-\exp\Big(-{(r-R_{\rm out})^2 \over r_s^2 }\Big)\Big]^5
                    \nonumber \mbox{.}
\end{eqnarray}

\noindent
We chose $m=2$ and $m=3$ perturbations in studying the stability properties
of our models.

Figure \ref{fig2}
plots the development of the global amplitudes for $m=1,2,3$, and
$m=4$ spiral modes in a stellar disk seeded by an $m=2$ perturbation of
the form (20). The global amplitudes defined by the expression
\begin{equation}
   A_{m}={\vert \int_{0}^{2 \pi}\int_{R_{\rm in}}^{R_{\rm out}}
       \sigma(r,\phi)dr
       e^{-im\phi}d\phi \vert \over{\int_{0}^{2 \pi}
       \int_{R_{\rm in}}^{R_{\rm out}}
   \sigma(r,\phi)dr d\phi}}
   \label{eq21}
\end{equation}
illustrate the overall dynamics of the particular global mode.

Figure \ref{fig2} shows a slow exponential growth of the $m=2$ mode. 
The $m=1$, $m=3$ and $m=4$ armed spirals grow, too, but
due to the initial conditions, the $m=2$ global mode outstrips
the other competitor modes during the whole computation.
However, even at the late stages of evolution the amplitude of 
spiral perturbations
is less then half percent, and the spiral pattern does not emerge from the
background. The slow development of the perturbation is best seen on the
sequence of the snapshots shown in Fig.\ \ref{fig3} illustrating the 
contour plots of the perturbed radial velocity. (The orbital periods at the
inner and outer boundary are 0.59 and 4.16, respectively.)

\begin{figure}[ptbh]
   \psfig{figure=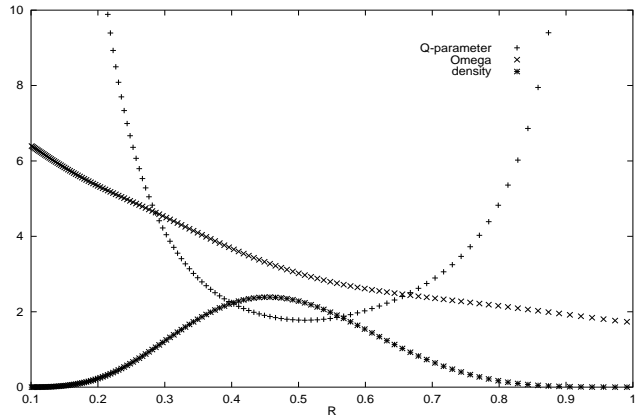,height=5.5cm,width=7.5cm,angle=270,bbllx=70pt,bblly=40pt,bburx=560pt,bbury=670pt}
   \caption[ ]{
          Radial dependence of the angular velocity $\Omega$,
          the Toomre-parameter $Q$, and the equilibrium density $\sigma$
          for a purely stellar disk ($\gamma=2$) with the Gaussian surface
          density distribution.
       \label{fig4}}
\end{figure}

Similar behavior was observed in the stellar disk with the Gaussian 
surface density distribution (\ref{eq18}) and the rotation curve (\ref{eq19}). 
Fig.\ \ref{fig4} shows the equilibrium properties of this disk which is rather
stable with a minimum value of Toomre's Q-parameter equal to 1.78 at a
disk radius of $R=0.5$.

\begin{figure}[ptbh]
   \psfig{figure=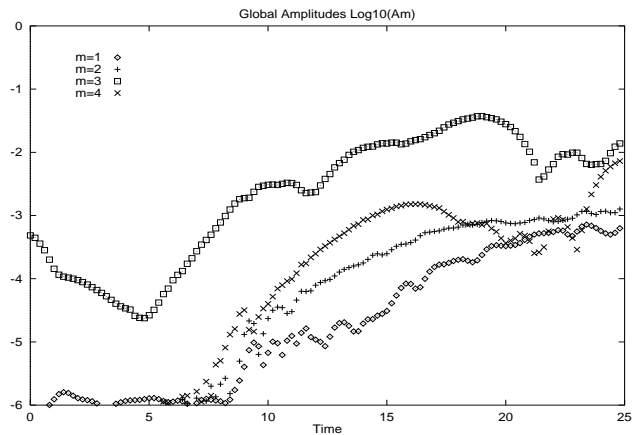,height=5.5cm,width=7.5cm,angle=270,bbllx=70pt,bblly=100pt,bburx=560pt,bbury=700pt}
   \caption[ ]{
          Temporal evolution of the global amplitudes $\log(A_m)$
          ($m=1,2,3,4$) for a purely stellar disk ($\gamma=2$)
          with the Gaussian surface density profile.
   \label{fig5}}
\end{figure}

This disk was seeded with a three-armed
perturbation of the form (\ref{eq20}). 
As it is seen from Fig.\ \ref{fig5}, the disk develops 
a set of slowly growing modes. The behavior of the global amplitudes is 
somewhat similar to the previous case. (The orbital periods at the
inner and outer boundary amount here to 0.97 and 3.65, respectively.)
The $m=3$ spiral mode has higher 
amplitude compared to the other competitors, and has a tendency to be 
saturated at the level $\log(A_3) \approx -2$. The perturbation, however, 
does not have any properties of a regular spiral pattern (Fig.\ \ref{fig6}). 
Even at the late stages of the evolution the contours remain quite
patchy, and do not resemble the regular spiral pattern.

\begin{figure*}[ptbh]
   \centerline{
   \psfig{figure=fig6.ps,height=21.cm,width=14.8cm,bbllx=0pt,bblly=0pt,bburx=550pt,bbury=750pt}
   }
   \caption[ ]{
          Contour maps of the logarithmic surface density 
          (${\rm M}_\odot / {\rm pc}^2$) 
          for a purely stellar Gaussian disk ($\gamma=2$) 
          at different times: $t=0$ (upper left), $t=5$ (upper right), 
   \label{fig6}}
\end{figure*}

In the next section we will discuss how the morphological properties
of the global modes are affected by the $\it change$ of the equation
of state which is a necessary consequence of a phase transformation in a
star-forming disk.


\section{Global Modes in a Multi-Component Disk}

\begin{figure}[ptbh]
   \psfig{figure=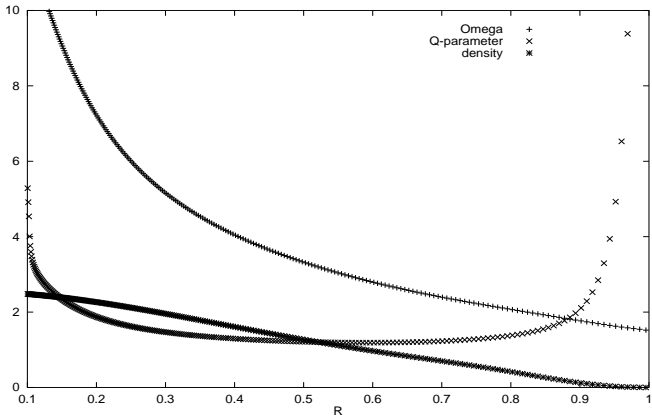,height=5.5cm,width=7.5cm,angle=270,bbllx=70pt,bblly=40pt,bburx=560pt,bbury=650pt}
   \caption[ ]{
          Radial dependence of the angular velocity $\Omega$, 
          density distribution and the Toomre-parameter $Q$
          for the gaseous component in a multi-component disk
          with the exponential surface density distribution.
   \label{fig7}}
\end{figure}

To compare the behavior of a multi-component disk with the 
dynamics of a corresponding one-component system, 
the surface densities of all phases  
were initially distributed in accordance with equations (\ref{eq16}) or 
(\ref{eq18}). In both cases we choose the initial mass of the gaseous 
component equal to the mass of the one-component system. 
Masses of the ``admixture'', i.e.\ stars and remnants, were initially set 
to 0.01 each which is about two percent of the initial mass of the
gaseous component. All components were set at the beginning into
centrifugal equilibrium, with the circular rotation supported by the  
pressure gradients of the components, the total gravitational 
field of the three components and the gravity of the external halo.
Fig.\ \ref{fig7} shows the rotation curve and the
epicycle frequency of the gaseous component for the initial exponential
surface density distributions. The rotation curve of the gaseous component 
is very similar to the equilibrium profiles of the stellar disk discussed 
in the previous section (Fig.\ \ref{fig1}). However, the gaseous component 
is less stable compared to the purely stellar disk, and the broad trough 
of Toomre's Q-parameter lies below the corresponding Q-distribution of the
stellar disk.

\begin{figure}[ptbh]
   \psfig{figure=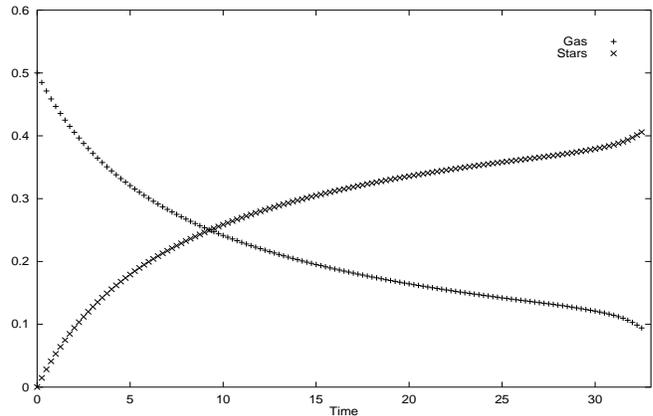,height=5.5cm,width=7.5cm,angle=270,bbllx=70pt,bblly=40pt,bburx=560pt,bbury=650pt}
   \caption[ ]{
          Temporal evolution of the masses of gaseous and the stellar component
          in a multi-component disk with an exponential 
          density distribution.
   \label{fig8}}
\end{figure}

Despite the centrifugal balance, the components are not in equilibrium. Mass 
transformations given by the right-hand-sides of the continuity
equations (\ref{eq6})-(\ref{eq8}) 
change the densities and total masses of the components, 
and the system evolves even without initial perturbations.
Fig.\ \ref{fig8} illustrates such mass transformation. At the beginning,  
the mass of the system is contained in the gaseous phase, 
and by the end of the computation
about 90\% of the gas has been converted into stellar remnants or long-lived
low mass stars. The fraction 
of mass contained in massive stars drops from the initial value to 
0.1\% at the end of the simulation.

\begin{figure}[ptbh]
   \psfig{figure=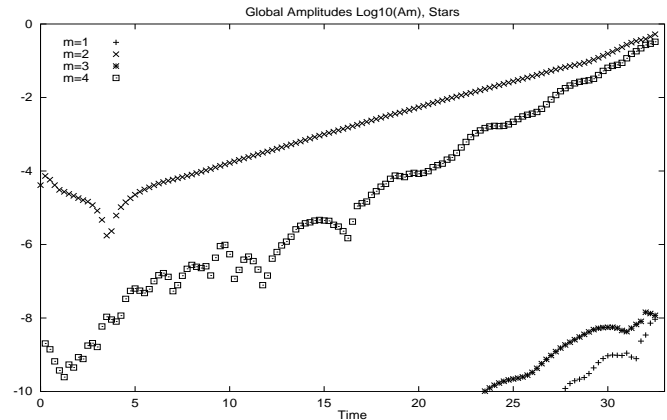,height=5.5cm,width=7.5cm,angle=270,bbllx=70pt,bblly=50pt,bburx=560pt,bbury=650pt}
   \caption[ ]{
          Temporal evolution of the global amplitudes $\log(A_m)$
          ($m=1,2,3,4$) for stellar component of the multi-phase disk
          with an exponential density distribution.
       \label{fig9}}
\end{figure}

With the $m=2$ perturbation given by equation (\ref{eq20}), 
all three phases develop a two-armed spiral pattern. 
Fig.\ \ref{fig9} shows the time dependence of the 
global amplitudes for the $m=1$,$2$,$3$ and $4$-armed global modes 
growing in the stellar component. 
Again, the $m=2$ global mode prevails over its competitors,
and compared to the purely stellar disk (Fig.\ \ref{fig2}) 
it grows about two orders of magnitude faster developing 
a nonlinear spiral pattern. 

Figures \ref{fig10} and \ref{fig11} show the time sequence of the contour 
plots of the density distributions of the stellar and gaseous components 
of a disk with an exponential density profile. Initially, the density 
distribution evolves similarly to the dynamics of the one-component disk, 
but the subsequent behavior is different. A comparison of Fig.\ \ref{fig10} 
with the simulations in a one-component disk 
clearly shows that spirals are better developed in a multi-component
disk. 

\begin{figure*}[ptbh]
   \centerline{
   \psfig{figure=fig10.ps,height=21.cm,width=14.8cm,bbllx=0pt,bblly=0pt,bburx=550pt,bbury=750pt}
   }
   \caption[ ]{
          Contour maps of the logarithmic surface density 
          (${\rm M}_\odot/{\rm pc}^2$) of the stars 
          for the multi-phase exponential disk
          at different times: $t=0$ (upper left), $t=5$ (upper right), 
          $t=10,15,20,25$. The contours are logarithmic-equally spaced.
          The equilibrium model is perturbed by an $m=2$-mode.
       \label{fig10}}
\end{figure*}

\begin{figure*}[ptbh]
   \centerline{
   \psfig{figure=fig11.ps,height=21.cm,width=14.8cm,bbllx=0pt,bblly=0pt,bburx=550pt,bbury=750pt}
   }
   \caption[ ]{
          Contour maps of the logarithmic surface density 
          (${\rm M}_\odot/{\rm pc}^2$) of the gas
          for the multi-phase exponential disk
          at different times: $t=0$ (upper left), $t=5$ (upper right), 
          $t=10,15,20,25$. The contours are logarithmic-equally spaced.
          The equilibrium model is perturbed by an $m=2$-mode.
       \label{fig11}}
\end{figure*}

\begin{figure}[ptbh]
   \psfig{figure=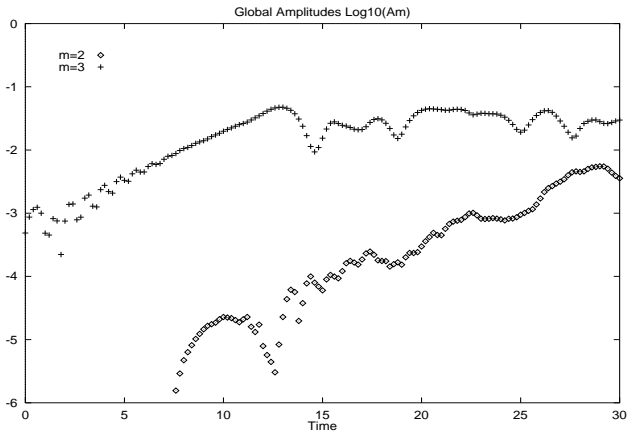,height=5.5cm,width=7.5cm,angle=270,bbllx=70pt,bblly=100pt,bburx=560pt,bbury=700pt}
   \caption[ ]{
          Temporal evolution of the global amplitudes $\log(A_m)$
          of the $m=2$ and $m=3$ modes
          for the stellar component of the multi-phase disk 
          with the Gaussian density profile.
   \label{fig12}}
\end{figure}

\begin{figure}[ptbh]
   \psfig{figure=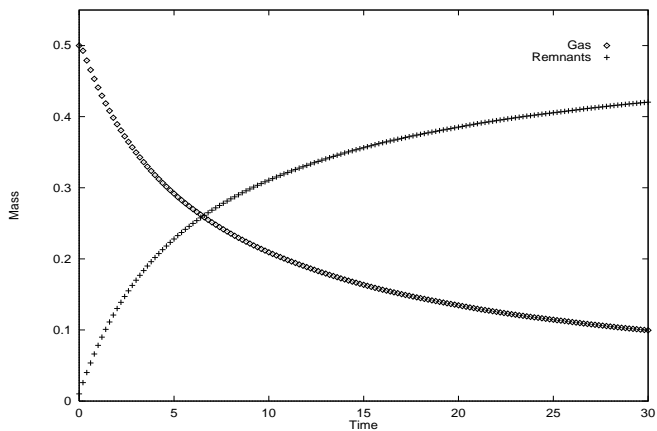,height=5.5cm,width=7.5cm,angle=270,bbllx=70pt,bblly=40pt,bburx=560pt,bbury=650pt}
   \caption[ ]{
          Mass transformations between the gaseous and stellar components
          in the multi-phase disk with the Gaussian density
          distribution.
   \label{fig13}}
\end{figure}

A similar behavior was observed for the disk with the Gaussian surface
density profile which was seeded by the $m=3$ perturbation. 
 Fig.\ \ref{fig12} shows the growth of the
global amplitude in the stellar component of the multi-phase disk
accompanied by the mass transformations illustrated on Fig.\ \ref{fig13}.
Again, a comparison with Fig.\ \ref{fig5} demonstrates, that the growth
rate of the $m=3$ spiral perturbation of the multi-phase disk is 
an order of magnitude larger than that of the one-component system.

The exponential growth of the
$m=3$ perturbation is changed by the lingering saturation phase
occurring at the amplitude level $\log(A_3) \approx -0.1$.  
This nonlinear saturation of global modes is known for one-component
simulations (Laughlin \& R\'o\.zyczka \cite{laughlin96}, Laughlin et al.\  
\cite{laughlin97}, \cite{laughlin98}). Our result shows that
the nonlinear saturation of exponentially growing global modes is
a common phenomenon which occurs as well in the multi-phase gravitating disks
experiencing phase transitions.

The three-armed nature of the perturbations in the multi-component  
disk is perfectly illustrated in Figs.\ \ref{fig14} and \ref{fig15} 
which show the contour map of the surface density and the 
radial velocity of the gaseous phase in a multi-component disk. 
We note that at late phases of the evolution the gas
contains about 8\% of the total disk mass, but nevertheless
it is still a good tracer of the spiral structure. Moreover, our
simulations allow to conclude, that the gas ''helps'' to develop spiral arms.  
A comparison of the density distribution of the collisionless remnants
phase in a multi-component disk (Fig.\ \ref{fig16}) with perturbations in a 
purely stellar disk (Fig.\ \ref{fig6}) supports this conclusion:
The density perturbation in remnants is more organized, and clearly depicts
a three-armed spiral.

\begin{figure*}[ptbh]
   \centerline{
   \psfig{figure=fig14.ps,height=21.cm,width=14.8cm,bbllx=0pt,bblly=0pt,bburx=550pt,bbury=750pt}
   }
   \caption[ ]{
          Contour maps of the logarithmic surface density 
          (${\rm M}_\odot/{\rm pc}^2$) of the gas
          for the multi-phase Gaussian disk
          at different times: $t=0$ (upper left), $t=5$ (upper right), 
          $t=10,15,20,25$. The contours are logarithmic-equally spaced.
          The equilibrium model is perturbed by an $m=3$-mode.
       \label{fig14}}
\end{figure*}

\begin{figure*}[ptbh]
   \centerline{
   \psfig{figure=fig15.ps,height=21.cm,width=14.8cm,bbllx=0pt,bblly=0pt,bburx=550pt,bbury=750pt}
   }
   \caption[ ]{
          Contour maps of the radial velocity (in kms$^{-1}$) 
          of the gaseous component in the Gaussian disk 
          at different times: $t=0$ (upper left), $t=5$ (upper right),
          $t=10,15,20,25$. The contours give 30\%, 50\%, 70\% and 90\% of
          the maximum velocity in each diagram. The dotted lines correspond
          to negative velocities, whereas the solid lines give positive
          velocities. The zero-velocity contour is shown with a dashed
          line.
          The equilibrium model is perturbed by an $m=3$-mode.
       \label{fig15}}
\end{figure*}

\begin{figure*}[ptbh]
   \centerline{
   \psfig{figure=fig16.ps,height=21.cm,width=14.8cm,bbllx=0pt,bblly=0pt,bburx=550pt,bbury=750pt}
   }
   \caption[ ]{
          Contour maps of the logarithmic surface density 
          (${\rm M}_\odot/{\rm pc}^2$) of the stellar remnant component
          for the multi-phase Gaussian disk
          at different times: $t=0$ (upper left), $t=5$ (upper right), 
          $t=10,15,20,25$. The contours are logarithmic-equally spaced.
          The equilibrium model is perturbed by an $m=3$-mode.
       \label{fig16}}
\end{figure*}

The destabilizing role of the cold gaseous component was studied for the linear
regime by various authors. Local analysis performed by Lin and Shu 
(\cite{lin66}), Lynden-Bell (\cite{lyndenbell67}), Miller et al.\ 
(\cite{miller70}), Quirk (\cite{quirk71}), Jog \& Solomon (\cite{jog84}),
Sellwood \& Carlberg (\cite{sellwood84}), and semi-analytical global 
modal analysis of Bertin \& Romeo (\cite{bertin88}) demonstrated 
that a small admixture of gas
in a stellar self-gravitating disk may considerably destabilize the system.
Our simulations are in agreement with this conclusion, and illustrate,
how the spiral structure behaves on a nonlinear stage.
Spiral structure remains well developed and is self-sustained after 
a rapid star-formation process in a gaseous disk when most of the gas 
is transformed into a ''remnants'' phase with higher velocity dispersion, 
and with a more rigid equation of state.


\section{Conclusions}

In this paper we modeled the generation of spiral structure
in a multi-phase, star-forming disk. The gaseous and stellar phases 
interact as well by their common gravity as
by mutual phase transitions due to star formation and stellar death.
The main results of our paper can be summarized as follows:

1. The multi-phase disks which 
   undergo a gas--star  phase transformations are
   unstable with respect to nonaxial perturbations. The global spiral modes
   grow exponentially and saturate in a way, similar to that
   found in the one-component case.

2. The spiral pattern grows faster and saturates on a higher level
   compared to a one-component stellar disk with the same
   mass and rotation.
   The spiral mode remains well developed, and keeps its
   properties unaltered if most of the gas is transformed into the
   ''stellar'' component with a stiffer equation of state. 
   The cold gas phase remains a good tracer of the spiral structure
   even if the system contains about a few percent of the total mass 
   of the disk.

   These results demonstrate that the destabilizing role of the
   cold gas component previously known for two-\-com\-po\-nent systems with
   fixed background properties can be extended to multi-component
   star-forming disks which are allowed for rapidly changing 
   background properties.

Further research, however, should be performed with respect to 
self-gravitating multi-component disks. One obvious generalization is
necessitated by the simplified model of the interaction chain used in our
paper. Interactions between the different phases in a star-forming region
are regulated mainly by the time dependent balance between the three
components: gas, molecular clouds and stars (e.g.\ Theis et al.\
(\cite{theis92}), Shore \& Ferrini \cite{shore95}). Therefore, in a 
next step the gaseous phase should be split into clouds and an 
inter-cloud medium matching the chemo-dynamical approach of 
e.g.\ Samland et al.\ (\cite{samland97}). Another aspect is the
stochasticity provided by the star-formation in individual clouds.
Thus, in extension of Gerola \& Seiden's (\cite{gerola78}) analysis based
on cellular automata, a study of the interplay between large-scale 
structure formation (including self-gravity) and stochastic processes 
would be interesting.


\begin{acknowledgements}
  We thank Greg Laughlin for providing his code and for valuable
  insights into the van Leer advection scheme. We are also grateful
  to Gerd Hensler for comments and discussions on chemo-dynamical
  evolution of galaxies. This work was supported
  by Deutsche Forschungsgemeinschaft with grant 436RUS17/121/96S.
  VK thanks Prof.\ Miyama for his kind hospitality at NAO where part of this
  work was performed.
\end{acknowledgements}


{}

\newpage

\appendix
\section{Parameters for Numerical Simulations}

%
%
\begin{table}[h]
\caption{Parameters for the exponential disk {\protect \\}
       (cf.\ Eqs.\ (\ref{eq16}) and (\ref{eq17}))}
\begin{tabular}{llll}
  \hline \hline
  General parameters: \\
  Courant factor            & $ C_{f}      =  0.6 $  \\
  Bulge mass                & $M_B         =  0.4$ \\
  Bulge scale length        & $R_B         =  0.4$ \\
  Halo  mass                & $M_H         =  1.6$ \\
  Halo scale length         & $R_H         =  0.32$ \\
  Inner disk radius         & $R_{\rm in}  = 0.1$ \\
  Outer disk radius         & $R_{\rm out} = 1.0$ \\

  \hline

  Mass exchange parameters: \\
  Stellar lifetime                   & $\tau   = 0.075$ \\
  Fraction of massive stars          & $ \zeta = 0.12$ \\
  Fraction of mass returned to ISM   & $\eta   = 0.9$ \\
  SFR parameter                      & $C_2    = 0.1$ \\

  \hline
  Disk (general):  \\
                          & $d_1 = 1.1$ \\    
                          & $d_2 = 8.0$ \\    
                          & $d_3 = 16.0$ \\    
                          & $r_s = 0.316$ \\    

  \hline
  Gaseous disk:  \\
  Initial mass             & $ M_{g}   = 0.22$  \\
  Polytropic index         & $\gamma_g = 1.67$ \\
  Popytropic coefficient   & $K_g      = 0.005$ \\

  \hline

  Stellar disk:  \\
  Initial mass             & $ M_{s}   = 0.001$  \\
  Polytropic index         & $\gamma_s = 2.0$ \\
  Popytropic coefficient   & $K_s      = 0.01$ \\

  \hline

  Remnants disk:  \\
  Initial mass             & $ M_{r}   = 0.001$  \\
  Polytropic index         & $\gamma_r = 2.0$ \\
  Popytropic coefficient   & $K_r      = 0.01$ \\
  \hline \hline
\end{tabular}
\end{table}
%
%

\begin{table}
\caption{Parameters for the Gaussian disk {\protect \\}
           (cf.\ Eqs.\ (\ref{eq18}) and (\ref{eq19}))}
\begin{tabular}{llll}
  \hline \hline
  General parameters: \\
  Courant factor            & $ C_{f}      =  0.5 $  \\
  Bulge mass                & $M_B         =  0.15$ \\
  Bulge scale length        & $R_B         =  0.2$ \\
  Halo mass                 & $M_H         = 10.0$ \\
  Halo scale length         & $R_H         =  0.25$ \\
  Velocity parameter        & $v_{\infty}  = 1.5$ \\
  Inner disk radius         & $R_{\rm in}  = 0.1$ \\
  Outer disk radius         & $R_{\rm out} = 1.0$ \\

  \hline

  Mass exchange parameters: the same\\

  Disk (general):  \\
                          & $r_s    = 0.316$ \\    
                          & $R_0    = 0.45$ \\    
                          & $\omega = 0.05$ \\    
  \hline

  Gaseous disk:  \\
  Initial mass             & $ M_{g}   = 0.5$  \\
  Polytropic index         & $\gamma_g = 1.67$ \\
  Popytropic coefficient   & $K_g      = 0.04$ \\

  \hline

  Stellar disk:  \\
  Initial mass             & $ M_{s}   = 0.01$  \\
  Polytropic index         & $\gamma_s = 2.0$ \\
  Popytropic coefficient   & $K_s      = 0.08$ \\

  \hline

  Remnants disk:  \\
  Initial mass             & $ M_{r}   = 0.01$  \\
  Polytropic index         & $\gamma_r = 2.0$ \\
  Popytropic coefficient   & $K_r      = 0.08$ \\
  \hline \hline
\end{tabular}\\
\end{table}

\end{document}